\documentclass[aps,pra,reprint,groupedaddress]{revtex4-1}
\usepackage{graphicx}% Include figure files
\usepackage{bm}% bold math
\usepackage{amsmath}
\usepackage[colorlinks=true, citecolor=blue, urlcolor=blue]{hyperref}
\begin{document}
%\preprint{}
\title{Closed-loop three-level charged quantum battery} %Closed-contour charged quantum battery}


\author{Fu-Quan Dou}
%\thanks{}
\altaffiliation{doufq@nwnu.edu.cn}
\affiliation{College of Physics and Electronic Engineering, Northwest Normal University,
Lanzhou, 730070, China}
\author{Yuan-Jin Wang}
\affiliation{College of Physics and Electronic Engineering, Northwest Normal University,
Lanzhou, 730070, China}

\author{Jian-An Sun}
\affiliation{College of Physics and Electronic Engineering, Northwest Normal University,
Lanzhou, 730070, China}

%\date{\today}

\begin{abstract}
Quantum batteries are energy storage or extract devices in a quantum system. Here, we present a closed-loop quantum battery by utilizing a closed-loop three-state quantum system in which the population dynamics depends on the three control fields and associated phases. 
We investigate the charging process of the closed-loop three-level quantum battery. The charging performance is greatly improved due to existence of the third field in the system to form a closed-contour interaction. Through selecting an appropriate the third control field, the maximum average power can be increased, even far beyond the most ideal maximum power value of non-closed-loop three-level quantum battery (corresponding to the most powerful charging obtainable with minimum quantum speed limit time and the maximum charging energy). We study the effect of global driving-field phase on the charging process and find the maximum extractable work (`ergotropy') and charging power vary periodically under different control field, with a period of $2\pi$. Possible experimental implementation in nitrogen-vacancy spin is discussed.
\end{abstract}

%\keywords{}

\maketitle

\section{Introduction}
Classical batteries on the basis of electrochemical principles are extremely useful to fulfil our daily life needs \cite{10.3389/fchem.2014.00079}. Currently, with the ever-increasing demand on the performance of energy storage devices, researchers have tried to exploit quantum phenomena to create a new class of powerful batteries which transcend conventional electrochemistry, i.e., quantum batteries \cite{Campaioli2018}.
%Ð޸Ĵ¦1
  This entirely new concept was first inroduced in \cite{PhysRevE.87.042123} and has become a very active research field \cite{caravelli2019random,PhysRevLett.111.240401,kamin2019nonmarkovian,PhysRevLett.124.130601,garcapintos2019fluctuations,Friis2017,alimuddin2020structure,juliafarre2018bounds,
  PhysRevB.99.205437,rossini2019quantum,rosa2019ultra,sen2019local,PhysRevB.99.035421,PhysRevLett.122.210601,liujpcc9b06373,zakavati2020bounds,quach2020using,
 santos2019controllable,PhysRevE.100.032107,Binder_2015,alimuddin2020structure,PhysRevLett.120.117702,PhysRevB.98.205423,PhysRevE.99.052106,zhang2018enhanced,
chen2019charging,PhysRevLett.122.047702,PhysRevLett.118.150601,PhysRevA.100.043833,PhysRevA.97.022106,PhysRevB.100.115142,PhysRevA.101.032115}. Quantum batteries are quantum device that can store or extract energy to perform work \cite{niedenzu2018quantum}. More specifically, a great number of researchers have recently addressed various aspects of quantum batteries, including work extraction \cite{caravelli2019random,PhysRevLett.111.240401,kamin2019nonmarkovian,PhysRevLett.124.130601,garcapintos2019fluctuations,Friis2017}, capacity \cite{juliafarre2018bounds}, role of entanglement and many-body interactions %rosa2019ultra,caravelli2019random,PhysRevB.100.115142},
  \cite{PhysRevB.99.205437,rossini2019quantum,rosa2019ultra,sen2019local,PhysRevB.99.035421} and environmental effects etc. \cite{PhysRevLett.122.210601,liujpcc9b06373,zakavati2020bounds,quach2020using}.
  %The system consists of $N$ quantum cells, that is, $N$ independent or entangled subsystems.
The maximum work that can be extracted from the quantum battery compatible with quantum mechanics is called ergotropy \cite{Allahverdyan_2004,santos2019controllable,PhysRevE.100.032107,Binder_2015,alimuddin2020structure}. Generally, the more energy stored, the higher the charging power, the better the battery performance. %Quantum batteries were first introduced in [8], followed by many articles [9?C14] regarding enhancement of charging power [15?C17], work extraction [18, 19] and advantage in multiple usage of the battery incorporated with entanglement [20?C22].  More specifically,a great number of researchers have recently addressed various aspects of quantum batteries
		
Up to present, most researches on quantum cells of quantum batteries focus on two-level systems \cite{PhysRevLett.120.117702,PhysRevB.98.205423,PhysRevE.99.052106,zhang2018enhanced,
chen2019charging,PhysRevLett.122.047702,PhysRevLett.118.150601,PhysRevA.100.043833} and spin chains \cite{PhysRevB.99.205437,PhysRevA.97.022106,PhysRevB.100.115142,PhysRevA.101.032115,rossini2019quantum,rosa2019ultra,juliafarre2018bounds}. For example, the collective charging scheme involves the concept of a Dicke quantum battery which consists of $N$ two-level systems, interacting with a photonic mode in a cavity to charge, and resulting in a quantum advantage in the charging power of a factor $\sqrt{N}$ \cite{PhysRevLett.120.117702}. Indeed, quantum mechanics can lead to an enhancement in the charging power when $N$ quantum cells are charged collectively \cite{PhysRevLett.118.150601}. The correlation-induced suppression of ergotropy is a characteristic of coherence on quantum batteries made of two-level systems and the disadvantage can be mitigated by considering the coherent optical state or modulating $N$ to approach infinity \cite{PhysRevLett.122.047702}. The spin-spin interactions between one-dimensional spin chain can yield an advantage in charging power, which comes from a mean-field interaction and relies on intrinsic interaction between quantum batteries \cite{PhysRevA.97.022106}. A quantum battery based on a disordered quantum Ising chain is characterized by high extractable work at low entanglement and suppression of energy fluctuation by interaction \cite{PhysRevB.100.115142}.

Three-level system, where two of the three available transitions
are coherently driven, is also an elementary building block of many quantum systems. It is widely used in light storage \cite{PhysRevLett.86.783}, atomic clock frequency standards \cite{vanier2005atomic} and coherent quantum control \cite{RevModPhys.79.53}, ranging from ultracold atoms \cite{Badshah_2019}, trapped ions \cite{PhysRevA.98.013423} to superconducting circuits \cite{inomata2016single} and Nitrogen-vacancy (NV) center \cite{PhysRevB.94.134107}. 
One of the advantages of three-level system over a two-level one is the additional controllability offered by the coupling field. In recent years, an important research topic in three-level system is that coherent driving of the third available transition forms a closed-contour interaction (the so-called closed-loop three-level system), which yields fundamentally new phenomena, including phase-controlled coherent population trapping and phase-controlled coherent population dynamics \cite{barfuss2018phase}.  The closed-loop interaction are used in detection and separation of chiral molecules \cite{PhysRevLett.122.173202}, coherent manipulation of a single spin \cite{barfuss2018phase} and adiabatic population transfer of a superconducting transmon circuit \cite{ineneaau5999}. It is very desirable to take the three-level system as the constituent unit of the quantum battery. Very recently, a three-level system is used to constitute a quantum battery and a stable charging process is realized by employing stimulated Raman adiabatic passage (STIRAP) technique \cite{PhysRevE.100.032107}. The three-level quantum battery allows one to avoid the spontaneous discharging regime. % When constructing quantum batteries by three-level systems, the can be used to avoid the spontaneous discharging regime \cite{PhysRevE.100.032107}. A natural question is what would be the impact of applying a closed-loop three-level system to the design of a quantum battery
Then a natural and interesting question is what would happen to the performance of a quantum battery if a closed-loop three-level system is applied to the design of a quantum battery.

In this paper, we consider a quantum battery for a closed-loop three-level system driven by three laser fields. We design the closed-loop three-level quantum battery model and study the charging dynamics, including charging energy and power. %Now we speculate that our battery have more charging power, after all, there are two different paths to connect the uncharged ground state to the fully charged active state.	
The rest of paper is organized as follows. In Sec. \uppercase\expandafter{\ref{2}} we begin with some basic concepts of quantum battery, and introduce the Hamiltonian of the system. Then we study the dynamic characteristics of quantum batteries in Sec. \uppercase\expandafter{\ref{3}}. A feasible scheme for realizing three-level quantum battery is described in Sec. \uppercase\expandafter{\ref{4}}. Finally, we have a summary in Sec. \uppercase\expandafter{\ref{5}}.
\section{\label{2} Model}
\begin{figure}
 	\includegraphics[width=0.4\textwidth]{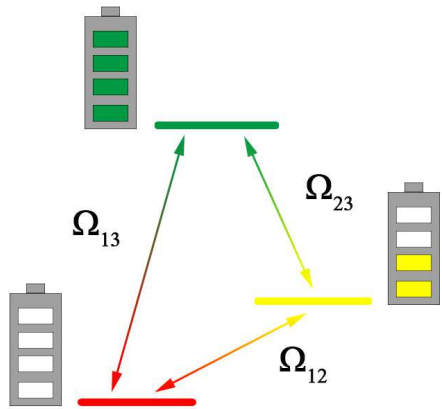}%
	 \caption{Visualisation of the concept of a closed-loop three-level quantum battery. Three discrete energy states represent different states of the battery and are charged through three laser fields: $\Omega_{12}, \Omega_{23}$  and $ \Omega_{13}$.  The spaces in the battery show the degree of charge, and the spaces gradually fill up from bare to fully charged batteries. When the three-level system is in the ground state (red), it is equivalent to a bare battery. We regard the intermediate state (yellow) as a partially charged battery. It represents a fully charged battery at the maximum excited state (green).}
	 \label{model}
\end{figure}
Without loss of generality, we assume a nondegenerate quantum battery by a %local and free
Hamiltonian \cite{Allahverdyan_2004},
\begin{equation}
	H_{0}=\sum_{n}^{d} \varepsilon_{n}\left|\varepsilon_{n}\right\rangle\left\langle\varepsilon_{n}\right|,  \quad     \varepsilon_{1}<\varepsilon_{2}<\cdots<\varepsilon_{d}.
\end{equation}
Our major objective is to analyze the performance of three-level quantum battery, so that $d=3$. We sketch our proposal in Fig. \ref{model}. At the initial moment, the system is prepared in the ground state, representing a depleted battery. To drive the system and promote transitions between the
energy levels, one utilizes auxiliary fields, i.e., suddenly switching on a transitional Hamiltonian $H_{1}$. We seek to inject as much energy into the quantum batteries as possible during the charging time $\tau$. The full Hamiltonian which describes the dynamics of the battery can be written as
	\begin{equation}
H(t)=H_{0}+\lambda(t)H_{\mathrm{1}}(t),
	\end{equation}
where $\lambda(t)$ is a dimensionless parameter, whose explicit dependence on time $t$ manifests the external control exerted on the system. For the sake of definiteness, we assume $\lambda(t)$ equal to one for $0<t<\tau$ and zero elsewhere. The character of the Hamiltonian $H_{\mathrm{1}}(t)$ is equivalent to a quantum charger, and $\lambda(t)$ guarantees it charge the battery only at $t \in[0,\tau]$.

The transitional Hamiltonian $H_{1}$ reads \cite{barfuss2018phase}
%\begin{eqnarray}
\begin{equation}
\begin{split}
&H_{\mathrm{1}}(t)=\\
&\hbar\left[\begin{array}{ccc}{0} & {\Omega_{12}(t)e^{-i (\omega_{12}t+\phi_1)}} & {\Omega_{13}(t) e^{-i (\omega_{13}t+\phi_3)}} \\ {\Omega_{12}(t)e^{i (\omega_{12}t+\phi_1)}} & {0} & {\Omega_{23}(t)e^{-i (\omega_{23}t+\phi_2)}} \\ {\Omega_{13}(t) e^{i (\omega_{13}t+\phi_3)}} & {\Omega_{23}(t)e^{i (\omega_{23}t+\phi_2)}} & {0}\end{array}\right]
\end{split}
\end{equation}
%\end{eqnarray}	
Here $\hbar$ is the reduced Planck constant. $\Omega_{12}, \Omega_{23}$  and $ \Omega_{13}$ are amplitudes %(Rabi frequencies)
of three driving fields (see Fig. \ref{model}). $\omega_{12}, \omega_{23}, \omega_{13}$ and $\phi_1, \phi_2, \phi_3$ are frequencies and phases of three driving fields, respectively. We assume they are real and positive.

The Hamiltonian $H_{0}$ plays a crucial role in how much energy a quantum battery stores. The total energy in the battery at time $t$ is implicitly
	\begin{equation}
E(t)=\operatorname{Tr}\left\{H_{0} \rho(t)\right\},
	\end{equation}
with $\rho=\sum_{n} r_{n}\left|r_{n}\right\rangle\left\langle r_{n}\right|$	being the density matrix of system \cite{Binder_2015}. And the time evolution of quantum state is according to the Liouville-von Neumann equation \cite{PhysRevE.87.042123}
	\begin{equation}
\dot{\rho}(t)=\frac{1}{i \hbar}\left[H(t), \rho(t)\right].
	\end{equation}	

Indeed, we explore the dynamics of the system charging in a time-dependent interaction picture and the Hamiltonian $H(t)$ can be written as \cite{PhysRevLett.122.090502}
%model the quantum dynamics via the following \cite{PhysRevLett.122.090502}	
	\begin{equation}
\label{HamInteractionP}
H_{\mathrm{int}}(t)=\hbar\left[\begin{array}{ccc}{0} & {\Omega_{12}(t)} & {\Omega_{13}(t) e^{i \phi}} \\ {\Omega_{12}(t)} & {0} & {\Omega_{23}(t)} \\ {\Omega_{13}(t) e^{-i \phi}} & {\Omega_{23}(t)} & {0}\end{array}\right],
\end{equation}
where we assumed that the driving fields are in resonance. %and satisfy $$
The %transitional Hamiltonian has a
global driving-field phase $\phi=\phi_1+\phi_2-\phi_3$, which %depends on the phases of the driving fields \cite{PhysRevLett.122.090502} and
 strongly influences the resulting dynamics \cite{barfuss2018phase}.	
	
The time evolution of quantum state is obtained from the equation
	 \begin{equation}
	\dot{\rho}_{\mathrm{int}}(t)=\frac{1}{i \hbar}\left[H_{\mathrm{int}}(t), \rho_{\mathrm{int}}(t)\right],
	\end{equation}	
with $\rho_{\text {int }}(t)=e^{i H_{0} t} \rho(t) e^{-i H_{0} t}$. In our charging protocol, we already assumed that three external fields mentioned above are in resonance with the levels of the battery. Therefore, the population in each energy level satisfies
	 \begin{equation}
	P_{n}=\operatorname{Tr}\left\{\hat{P}_{n} \rho_{\text {int }}(t)\right\}=\operatorname{Tr}\left\{\hat{P}_{n} \rho(t)\right\},
	\end{equation}	
where $\hat{P}_{n}$ represents the projector $\hat{P}_{n}=\left|\varepsilon_{n}\right\rangle\left\langle\varepsilon_{n}\right|$. Furthermore, $\operatorname{Tr}\left\{H_{0} \rho_{\text {int }}(t)\right\}=\operatorname{Tr}\left\{H_{0} \rho(t)\right\}$, the extractable work can be obtained from the difference,
	\begin{equation}
C(t)=\operatorname{Tr}\left\{H_{0} \rho_{\text {int }}(t)\right\}-\varepsilon_{1}.
	\end{equation}
Allowing the system to undergo adiabatic dynamics, the evolved state is $|\psi^{ad}(t)\rangle$, and the ergotropy is
\begin{eqnarray}
C(t)=&&\langle \psi^{ad}(t) |H_0| \psi^{ad}(t)\rangle-\langle\varepsilon_{1}|H_0|\varepsilon_{1}\rangle \nonumber\\
=&&\langle \psi^{ad}(t) |H_0| \psi^{ad}(t)\rangle-\varepsilon_{1}.
\end{eqnarray}		
Notice that if $\varepsilon_{1}=0$ then ergotropy coincides with the mean energy of $\rho_{\text{int}}$, i.e., $ C(t)=E(t)$. For the total charging time $\tau$, the corresponding work is $C(\tau)$
	\begin{equation}
P(\tau)=\frac{C(\tau)}{\tau}.
	\end{equation}
In Ref. \cite{PhysRevE.100.032107}, $\Omega_{13}=0$, the STIRAP protocol can avoid the oscillatory behavior and achieve a stable charging process. Consistent with the previous works \cite{PhysRevE.100.032107}, we take %regulate $\Omega_{12}(t)=\Omega_{0} f(t)$ and $\Omega_{23}(t)=\Omega_{0}[1-f(t)]$, where $\Omega_{0}$ is a constant and $f(t)=t/\tau$.
\begin{equation}
\Omega_{12}(t)=\Omega_{0} f(t), \quad \Omega_{23}(t)=\Omega_{0}[1-f(t)],
	\end{equation}
where $\Omega_{0}$ is a constant and $f(t)=t/\tau$. In what follows, we calculate and analyze the ergotropy and the charging power for different parameters shown in the interaction Hamiltonian $H_{\mathrm{int}}$.
\section{\label{3}Dynamics in closed-loop quantum batteries}%	
We now analyze the charging process focusing on the closed-loop quantum battery. We first consider a special case, i.e., the phase $\phi=\pi/2$. The corresponding eigenstates of the Hamiltonian (\ref{HamInteractionP}) are
\begin{eqnarray}
|E_{-}(t)\rangle&=&\frac{1}{\sqrt{2}}\left(\frac{\Omega_{12}(t)\Omega_{23}(t)}{\Omega(t) \Omega_1(t)}-i \frac{\Omega_{13}(t)}{\Omega_1(t)}\right)|\varepsilon_1\rangle\nonumber\\&&-\frac{1}{\sqrt{2}}\left(\frac{\Omega_{23}(t)}{\Omega_1(t)}-i \frac{\Omega_{12}(t)\Omega_{13}(t)}{\Omega(t) \Omega_1(t)}\right)|\varepsilon_2\rangle\nonumber\\&&+\frac{1}{\sqrt{2}} \frac{\Omega_{1}(t)}{\Omega(t)}|\varepsilon_3\rangle,\\
|E_{0}(t)\rangle&=&\frac{\Omega_{23}(t)}{\Omega(t)}|\varepsilon_1\rangle+i\frac{\Omega_{13}(t)}{\Omega(t)}|\varepsilon_2\rangle-\frac{\Omega_{12}(t)}{\Omega(t)}|\varepsilon_3\rangle,\\
|E_{+}(t)\rangle&=&\frac{1}{\sqrt{2}}\left(\frac{\Omega_{12}(t)\Omega_{23}(t)}{\Omega(t) \Omega_1(t)}+i \frac{\Omega_{13}(t)}{\Omega_1(t)}\right)|\varepsilon_1\rangle \nonumber\\&&+\frac{1}{\sqrt{2}} \left(\frac{\Omega_{23}(t)}{\Omega_1(t)}+i \frac{\Omega_{12}(t)\Omega_{13}(t)}{\Omega(t) \Omega_1(t)}\right)|\varepsilon_2\rangle\nonumber\\&&+\frac{1}{\sqrt{2}} \frac{\Omega_{1}(t)}{\Omega(t)}|\varepsilon_3\rangle,
\end{eqnarray}
with the eigenenergies $E_{0}(t)=0$ and $E_{\pm}(t)=\pm\hbar\Omega(t)$, where $\Omega^{2}(t)=\Omega^{2}_{12}(t)+\Omega^{2}_{23}(t)+\Omega^{2}_{13}(t)$ and $\Omega^{2}_1(t)=\Omega^{2}_{13}(t)+\Omega^{2}_{23}(t)$.

To achieve an efficient charged state, we employ the STIRAP technique \cite{RevModPhys.89.015006,PhysRevA.87.043631}  and assume the initial state of system $|\psi(0)\rangle=|E_0(t)\rangle=|\varepsilon_1\rangle$. Therefore, the initial values of the control fields satisfy $\Omega_{12}(0)=\Omega_{13}(0)=0$ and $\Omega_{23}(t)\neq0$. When the system undergoes adiabatic dynamics, the evolved state becomes $|\psi^{ad}(t)\rangle=|E_0(t)\rangle$. Then the ergotropy is
\begin{equation}\label{ergotropy}
C(t)=\frac{\Omega^2_{23}}{\Omega^2(t)}\varepsilon_1+\frac{\Omega^2_{13}}{\Omega^2(t)}\varepsilon_2+\frac{\Omega^2_{12}}{\Omega^2(t)}\varepsilon_3-\varepsilon_1.
\end{equation}
One find that the ergotropy depends on the final values for control fields $\Omega_{12}(t), \Omega_{23}(t)$ and $\Omega_{13}(t)$ at some cutoff time $\tau_{c}$ and can arrive the maximal value $C_{max}=\varepsilon_3-\varepsilon_1$ when $\Omega_{13}(\tau_c)=\Omega_{23}(\tau_c)=0$ and $\Omega_{12}(\tau_c)\neq0$. To the end we select the suitable control fields, such that the above boundary conditions are satisfied. In the following calculation we take $\varepsilon_{1}=0$, $\varepsilon_{2}=\hbar$, $\varepsilon_{3}=1.95\hbar$, respectively. The control fields $\Omega_{13}(t)$ is taken as
\begin{eqnarray}\label{Omega13}
\Omega_{13}(t)=&\Omega_0\sin(\pi t/\tau), \,\, \Omega_0(1-\cos(2\pi t/\tau)),\nonumber \\ &\Omega_0(1-\cos(2\pi t/\tau))^2.
\end{eqnarray}
\begin{figure}[htbp]
	\centering
	\centering
	\includegraphics[width=0.465\textwidth]{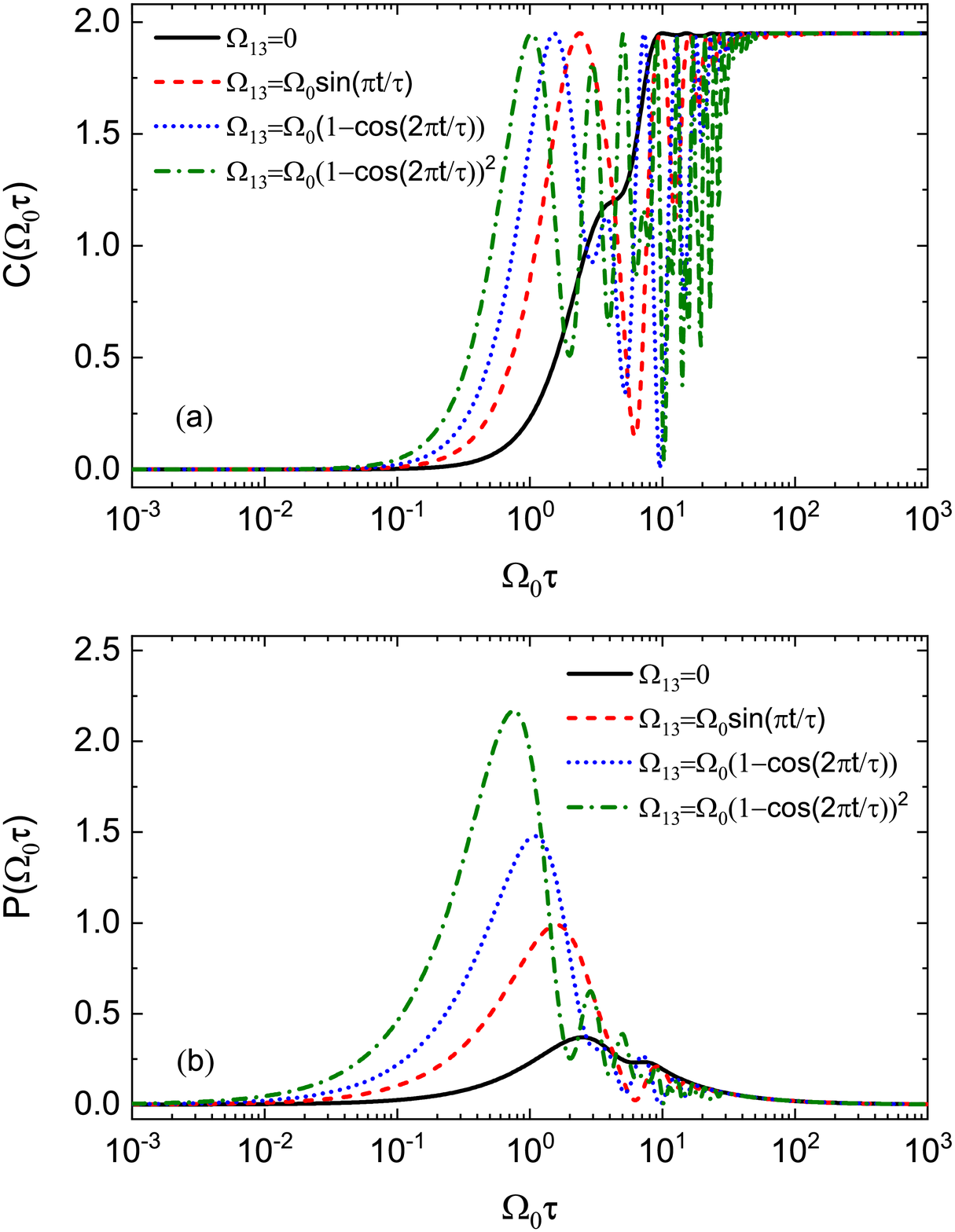}
\caption{(a) The dependence of the ergotropy on $\Omega_0\tau$ for $\Omega_{13}=\Omega_{0}\sin(\pi t/\tau)$ (red dashed), $\Omega_{13}=\Omega_{0}(1-\cos{(2\pi t/\tau)})$ (blue dotted) and $\Omega_{13}=\Omega_{0}(1-\cos{(2\pi t/\tau)})^2$ (olive dash-dotted). The solid black curve represents the non-closed case $\Omega_{13}=0$. (b) The average charging power as a function of $\Omega_0\tau$. Color coding and labeling is the same as in (a). All shown data have been computed by setting $\phi=\pi/2$.}
	\label{Fig.2}
\end{figure}	
	
The dependence of the ergotropy $C(\tau)$ and the average charging power $P(\tau)$ on $\tau$ is indicated in Fig. \ref{Fig.2} for several categories of $\Omega_{13}$. Here $\Omega_0 \tau$ is the dimensionless parameter. %The horizontal axis is a exponent with a base of $10$, ranging from $0.001$ to $1000$. %In contrast, no further rescale be applied to the results of numerical simulations $C(\tau)$ and $P(\tau)$.
 In order to analyze the advantages of closed-loop, we also plot the situation of non-closed loop (solid black line), corresponding to $\Omega_{13}=0$. Different from the non-closed loop case, the evolution process of ergotropy is divided into four windows along the $\tau$ dimension. For a fast evolution, the ergotropy is almost $0$ due to being far from the adiabatic limit at these timescales. With increasing $\tau$ the ergotropy grows monotonically to a maximum value, which achieves a fully charged state (corresponding to the maximum charging energy), then begins to oscillate, and finally reaches and stays at it's maximum value for large timescales. We also clearly see that, for different control fields, because of the different transfer time of all the population from the initial ground state to the maximally excited state, the minimal time to reach the maximum ergotropy for the first time is different. As a result, for fast protocol our batteries fail to charge (corresponding to small average power). However, as $\tau$ increases, the average charging power also increases until it reaches a maximum value at some point. Beyond this timescale, the average power will be less than this maximum value. %This means that the charging energy first arrive to it's the maximum value at the moment. %%One find that the maximum average charging power of closed-contour battery is greatly improved, which the value closes to $4$ times of the original non-closed loop battery for $\Omega_{13}=\Omega_{0}\sin(\pi t/\tau)$. When we select the control field $\Omega_{13}=\Omega_{0}(1-\cos{(2\pi t/\tau)})^n$ ($n=1,2,\cdots$), the value will continue to increase.%we prepared a additional curve (solid black line), which represents the situation of non-closed loop ($\Omega_{13}=0$) as a comparison. %As we increase ?? , in line with the adiabatic theorem [45,53,54], the maximum ergotropy grows, and we achieve a fully charged state when the STIRAP protocol is faithfully implemented. We clearly see that in the case of no decoherence the charged state is perfectly stable for ??  10/	0.	
%For fast protocol where our batteries have advantages in terms of energy storage and average charging power, and they reach the maximum energy storage and average charging power more quickly.

It is interesting to note that the maximum average charging power of closed-loop battery is greatly improved, even far beyond the ideal maximum power value of non-closed-loop three-level quantum battery, corresponding to the most powerful charging obtainable with minimum quantum speed limit time and the maximum charging energy \cite{PhysRevE.100.032107}. More detailed, the value is close to $4$ times of that for the original non-closed loop battery with $\Omega_{13}=\Omega_{0}\sin(\pi t/\tau)$. When we select the control field $\Omega_{13}=\Omega_{0}(1-\cos{(2\pi t/\tau)})^n$ ($n=1,2,\cdots$), the value will continue to increase and is close to $6$ times that for the original non-closed loop battery when $\Omega_{13}=\Omega_{0}(1-\cos{(2\pi t/\tau)})$ and more than $8$ times as high when $\Omega_{13}=\Omega_{0}(1-\cos{(2\pi t/\tau)})^2$.  Further study shows that the maximum average charging power can be increased by increasing the index $n$ of control field. Furthermore, compared with the ergotropy of the non-closed loop system at its maximum average charging power, the closed-loop batteries charged by three laser fields can store more ergotropy, i.e., the existence of the control field $\Omega_{13}$ can greatly improve the maximum average charging power and the extractable energy, thus accelerating the charging process. Therefore, for an optimized $\Omega_{13}$, the system can realize high efficient and stable charging process as long as we immediately turn off the $H_1$ after the moment of reaching the maximum average charging power.
\begin{figure}[htbp]
	\centering
	\centering
	\includegraphics[width=0.465\textwidth]{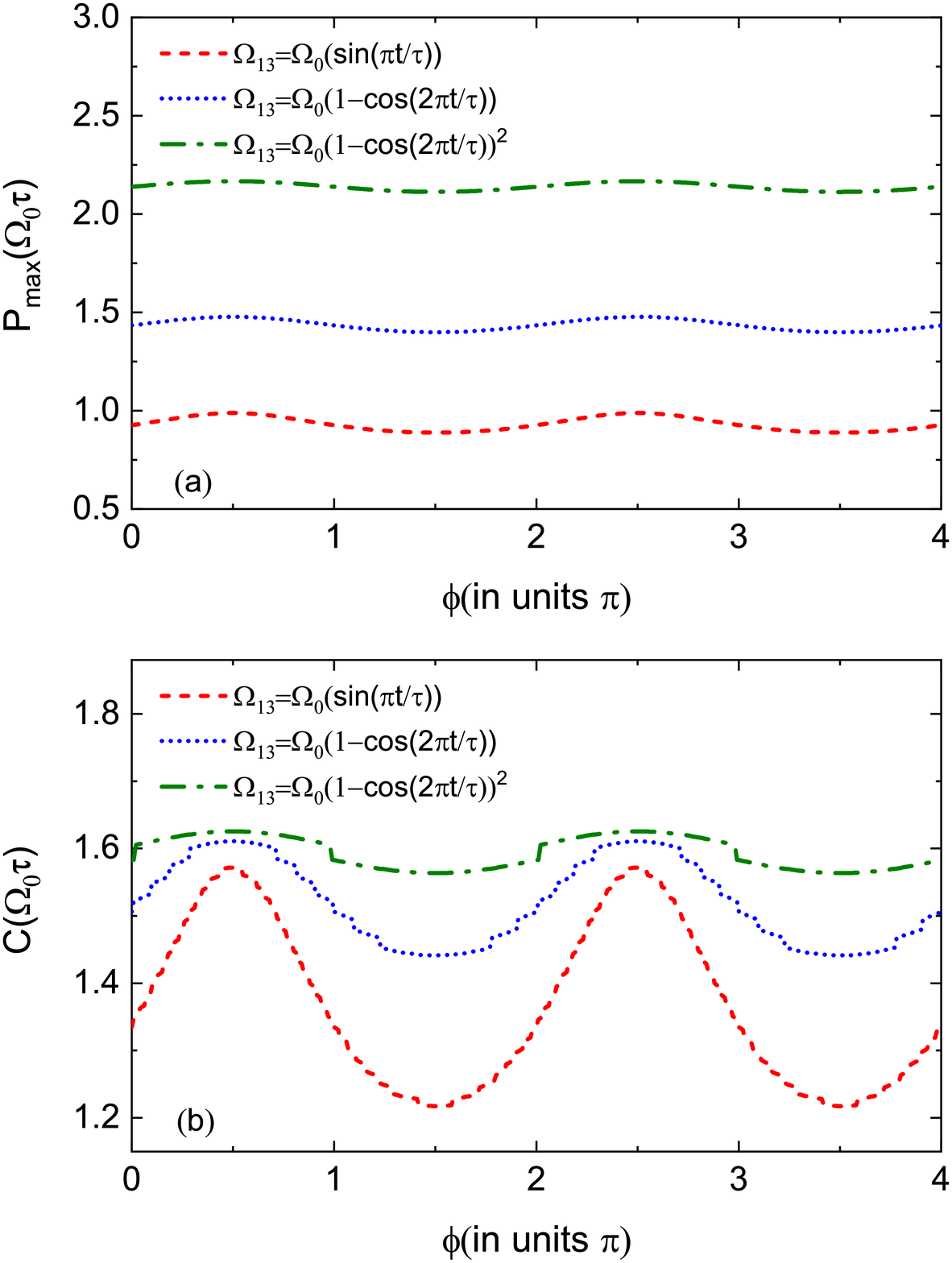}
	\caption{(a) The maximum charging power and (b) the corresponding energy as a function of  $\phi$ for $\Omega_{13}(t)=\Omega_{0}\sin(\pi t/\tau)$  (red dashed), $\Omega_{13}(t)=\Omega_{0}(1-\cos{(2\pi t/\tau)})$  (blue dotted) and $\Omega_{13}=\Omega_{0}(1-\cos{(2\pi t/\tau)})^2$ (olive dash-dotted). }
\label{Fig.3}
	\end{figure}

\begin{figure}[htbp]
	\centering
	\centering
	\includegraphics[width=0.465\textwidth]{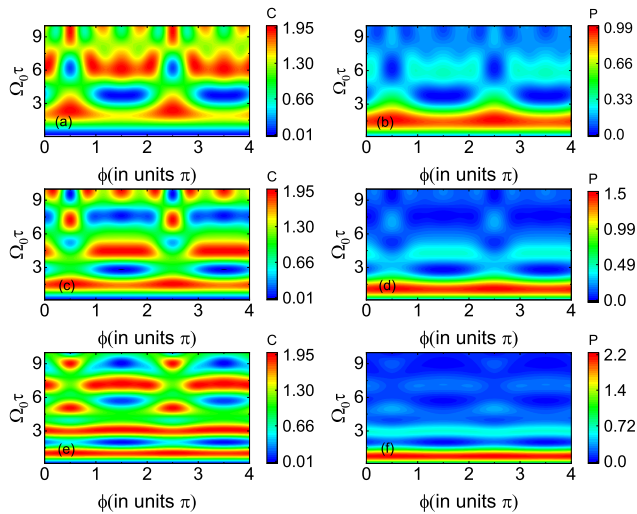}
	\caption{Contour plots of charging energy (left columns) and power (right columns) as the function of phase $\phi$ and charging time $\tau$ for different $\Omega_{13}$. (a) and (b) $\Omega_{13}=\Omega_{0}\sin(\pi t/\tau)$. (c) and (d) $\Omega_{13}=\Omega_{0}(1-\cos{(2\pi t/\tau)})$. (e) and (f) $\Omega_{13}=\Omega_{0}(1-\cos{(2\pi t/\tau)})^2$.
	 }
\label{Fig.4}
	\end{figure}

So far we only consider the case of the total driving phase $\phi=\pi/2$. To further demonstrate the effect of the phase on the charging process, we calculate the maximum average charging power and the charging energy at different phases. Fig. \ref{Fig.3} shows the maximum average charging power and the corresponding energy as a function of $\phi$ under the different control field $\Omega_{13}$. No matter what the control field is, the maximum charging power and energy have the same period $2\pi$ and reach their maximum value at $\phi=\pi/2$. As the index $n$ increases, the maximum average charging power and the corresponding energy increase and the amplitude of oscillation decreases. For a clearer and more comprehensive understanding of the effect of phase on the charging energy and the average charging power,
in Fig. \ref{Fig.4}, we display the charging energy and the average charging power as a function of both
phase $\phi$ and charging time $\tau$ for different control field $\Omega_{13}$. The blue zones correspond to low value whereas red areas indicate high value. The plot reveals main features of the charging process with phase and
charging time. It's apparent that phase plays a non-negligible role in charging process. The charging energy and the maximum average charging power are obviously periodic with $2\pi$ and can obtain the maximum value at the position $\phi=\pi/2$.

\section{\label{4}Quantum battery implement in nitrogen-vacancy spin}
There are several physical systems to implement the closed-loop three-level quantum battery such as trapped ion systems or superconducting circuit systems. Here we briefly describe a scheme which coherently drives the NV spin using a combination of time-varying magnetic and strain fields to implement a three-level quantum battery. The negatively charged NV centre in the diamond lattice forms an $S=1$ spin system. Under an appropriate rotating frame and the resonant case, the dynamics of the NV spin are described by the Hamiltonian (\ref{HamInteractionP}) \cite{barfuss2018phase,PhysRevLett.114.145502}. Conveniently, the initialization of the system can be realized by means of optical spin pumping under green laser excitation. Even at room temperatures, the spin of the NV can also be initialized easily. This character makes it become platforms for quantum information processing \cite{PhysRevB.94.134107}. The three eigenstates of the spin operater $\hat{S}_{z}$ are $|-1\rangle$, $|0\rangle$ and $|+1\rangle$, which correspond to $\left|\varepsilon_{1}\right\rangle$, $\left|\varepsilon_{2}\right\rangle$ and $\left|\varepsilon_{3}\right\rangle$ in our battery, respectively. Thus, one can prepare the system correspond to a bare quantum battery in $\left|\varepsilon_{1}\right\rangle$ at first, and then utilize a time-varying strain field to drive $|-1\rangle \leftrightarrow|+1\rangle$ transition and microwave magnetic fields to drive $|0\rangle \leftrightarrow|\pm 1\rangle$ transitions. The NV spin can be optically read out by virtue of its spin-dependent fluorescence. At last, the ergotropy and charging power of the three-level quantum battery can be obtained by uncomplicated calculation.

\section{\label{5}Conclution}
We have introduced the concept of a ''closed-loop three-level quantum battery'', which is a three-level system driven by three available transitions forming a closed-contour interaction. We show the performance of the quantum battery can be greatly improved by choosing an appropriate the third driving field. The closed-contour interaction makes the maximum average charging power can be greatly increased, even far beyond the most ideal maximum power value of non-closed-loop three-level quantum battery. In addition, the charging energy and power can reach the peak value at phase $\phi=\pi/2$ and vary with phase with a period of $2\pi$. Finally, we have briefly described the scheme of realizing closed-loop three-level quantum battery by a nitrogen-vacancy spin system.

 %a significant improvement in the maximum charging power with respect to non-closed-loop case. %two-fold enhanced scaling of . have advantages in the ergotropy and power within a certain charging time. When the charging time $\tau$ dose not exceed the critical value, the presence of a third driving field enables the battery to have more charge power than a non closed-loop system, and makes the maximum average charging power of our quantum battery close to or even more than twice, while storing more extractable energy at the maximum charge power. %However, once $\tau$ exceeds the threshold not only has no advantage over the charging power but also causes the ergotropy to oscillate.
 %In addition, the maximum average charging power varies with phase with a period of $2\pi$ and reaches a peak at $\phi=\pi/2$. Finally, we have briefly described the scheme of realizing closed-loop three-level quantum battery by a nitrogen-vacancy spin system.	
\begin{acknowledgments}
The work is supported by the National Natural Science Foundation of China (Grants No. 11665020).
\end{acknowledgments}
\bibliography{Reference}

\end{document}